\documentclass[doublecol]{epl2} 

\usepackage{amsmath} \usepackage{amssymb} \usepackage{cite} \usepackage{color}

\newcommand{\kk}{\mathbf{k}} 

\newcommand{\xx}{\mathbf{x}}
\newcommand{\yy}{\mathbf{y}}

\newcommand{\ee}{\mathrm{e}}
\newcommand{\dd}{\mathrm{d}}

\newcommand{\eq}[1]{(\ref{#1})}
\newcommand{\eqname}[1]{\label{#1}}

\title{Non-equilibrium quasi-condensates in reduced dimensions}

\author{A. Chiocchetta\inst{1,2} \and I. Carusotto\inst{3}}
\shortauthor{A. Chiocchetta \etal}
\institute{                    
  \inst{1} Dipartimento di Fisica, Universit\`a di Trento, via Sommarive 14, 38123 Povo, Italy\\
 \inst{2} SISSA - International School for Advanced Studies, via Bonomea 265, 34136 Trieste, Italy \\
  \inst{3} INO-CNR BEC Center and Dipartimento di Fisica, Universit\`a di Trento, via Sommarive 14, 38123 Povo, Italy\\
}

\pacs{71.36.+c}{Polaritons} 
\pacs{03.75.Kk}{Dynamic properties of condensates}
\pacs{05.70.Ln}{Nonequilibrium and irreversible thermodynamics}

\abstract{We develop a generic {phenomenological model to describe} the fluctuations on top of a non-equilibrium Bose-Einstein condensate. Analytic expressions are obtained for the momentum distribution of the non-condensed cloud and for the long-distance behavior of the spatial coherence {in the different dimensionalities}. Comparison of our predictions with available experimental data on condensates of exciton-polaritons and on surface-emitting planar laser devices is finally made.}

\begin{document}

\maketitle

Non-equilibrium phase transitions are among the most exciting topics of non-equilibrium statistical mechanics~\cite{Ruelle}. In the last decades they have been studied in a variety of different physical systems, of either classical~\cite{zia,mukamel,zia2} or quantum \cite{FRENS1990,DIEHL} nature. However, as compared to their equilibrium counterparts, much less is known about their general features, in particular for what concerns the critical behavior in the vicinity of the transition point. A most important class of non-equilibrium phase transitions take place in systems at the so-called non-equilibrium steady state (NESS) where a dynamical balance of driving and dissipation replaces the usual thermal equilibrium condition of standard equilibrium statistical mechanics.

Even though laser operation has played a central role in most developments of contemporary experimental atomic, molecular and optical physics, its potential as a workbench for non-equilibrium statistical mechanics studies has not been fully exploited yet. The interpretation of the laser threshold as a second-order phase transition dates back to the early 1970's with the pioneering works by DeGiorgio and Scully~\cite{digiorgio} and by Graham and Haken~\cite{haken70,Haken}: the order parameter of the transition is the amplitude $E$ of the electromagnetic field in the laser cavity, which gets a well defined phase by spontaneously breaking the $U(1)$ symmetry corresponding to phase rotations, $E\rightarrow E\,e^{i\theta}$. 

While most textbook discuss this analogy in the case of single-mode cavities only, the spontaneous symmetry breaking phenomenon is -rigorously speaking- restricted to spatially extended systems. Only in this case, it is in fact meaningful to take the long-distance limit involved in the Penrose-Onsager definition of an ordered state,
$\lim_{|\xx-\xx'|\to\infty} \langle \hat{E}^\dagger(\xx)\,\hat{E}(\xx') \rangle \neq 0$.
Most remarkable examples of spatially extended laser devices are the so-called vertical cavity surface emitting lasers (VCSELs), whose planar geometry makes a continuum of in-plane modes to be available to both lasing and fluctuations{: in the last years, VCSELs have attracted a strong technological interest from the point of view of nonlinear optics and of all-optical information processing~\cite{VCSEL}. Still, not much attention has been devoted to the statistical features of their operation: even though their importance for low-dimensional devices was recognized already in~\cite{haken70}, very little experimental work has been devoted to, e.g., the study of the microscopic processes limiting the spatial coherence of devices of different geometry and dimensionality. A pioneering work in this direction was reported in~\cite{Kapon}. }

{On the other hand, a great attention has been} devoted in the last years to the Bose-Einstein condensation (BEC) phase transition in gases of exciton-polaritons in semiconductor microcavities~\cite{KASP2006}. In the strong light-matter coupling regime, the cavity photon is strongly mixed to an excitonic transition in the cavity medium, giving rise to new long-lived bosonic excitation modes, the so-called exciton-polaritons{. For} strong enough pump intensities, the density {of the polariton gas} reaches the threshold for BEC and long-range spatial and temporal coherence appear. Recently, BEC was also observed in a photon gas in a macroscopic cavity filled with dye molecules~\cite{KLAERS2010}.

Even though {many} features of these BEC experiments are accurately described under the assumption that thermal equilibrium is reached within the photon/polariton gas on a time-scale shorter than the particle decay rate $\gamma$, several authors have pointed out intriguing consequences of the non-equilibrium nature of these optical BECs, whose steady-state is determined by a dynamical balance of pumping and particle losses: the Goldstone mode associated to the spontaneously broken $U(1)$ symmetry has a diffusive rather than sonic nature~\cite{WOUT2006,SZYM2006,WOUT2007} and the momentum-space shape of the condensate strongly depends on the pump spot geometry~\cite{RICH2005b,WOUT2008}.

{Starting from the fact that polariton BEC, photon BEC, and laser operation can all be described in terms of a spontaneously broken $U(1)$ symmetry, the purpose of this Letter is to develop a generic phenomenological model of {\em non-equilibrium Bose-Einstein condensation} that can be applied to all these systems: stimulated emission of photons by a population-inverted laser medium corresponds in the BEC case to the stimulated scattering of a thermal particle into the condensate mode.

As a first application, we apply this theory to study the long-distance behavior of the spatial coherence of (quasi)-condensates in different dimensionalities. Starting from the pioneering work~\cite{haken70}, this has been an active subject of research using stochastic field~\cite{CARU2005,WOUT2006} and Keldysh techniques~\cite{SZYM2006,SZYM2012,ROUM2012}: nowadays {there is a quite general consensus that}, independently of its microscopic details, the long-distance behaviour of fluctuations in the non-equilibrium steady-state of an interacting system matches the one of the corresponding equilibrium system at a finite temperature.

 The main result of this Letter is that this crucial conclusion can be recovered using an analytically tractable model based on a linearized stochastic Gross-Pitaevskii equation (SGPE). A main advantage of our model lies in its physical transparency and in the facility in which it can be extended to include new features: the final part of this Letter will be devoted to a brief discussion of the novel physics introduced by a frequency-dependence of the amplification baths.}

\section{Stochastic Gross-Pitaevskii Equation}
The basic idea underlying the SGPE is to start from the mean-field equation for the coherent condensate field and then to add a stochastic noise term to include quantum and/or classical fluctuations. 
Independently of the microscopic nature of the Bose field $\phi(\xx,t)$ (which can describe equally well the electromagnetic field of a VCSEL and the polariton field of a polariton condensate), its low-energy physics can be accurately described by a $d$-dimensional SGPE of the form~\cite{WOUT2007}: 
\begin{equation}\label{GGPE}
 i\,\dd\phi  =  \left[ \omega_0 -\frac{\hbar \nabla^2}{2m} + g|\phi|^2 + i\left( \frac{P_0}{1 + \frac{|\phi|^2}{n_s}} - \gamma \right) \right]\phi\,\dd t+\dd W, 
\end{equation}
where $m$ is the boson mass of the field, $\gamma$ is the damping rate, $P_0$ is the pumping rate and $n_s$ is the saturation density of the pumping mechanism. As usual, the validity of the SGPE model requires that the nonlinear coupling strength $g\geq 0$ describing repulsive contact interactions between the particles be weak enough for the gas to be dilute.

As usual in open systems, the presence of dissipative terms in the motion equations requires including corresponding stochastic terms $\dd W(\mathbf{x},t)$. For the sake of simplicity, we phenomenologically choose the simplest Ito form of a white complex Gaussian noise term with a random phase and spatially and temporally local correlations,
\begin{eqnarray}
 \left \langle \dd W(\mathbf{x},t)\dd W^*(\mathbf{x'},t)   \right \rangle & = & 2D_{\phi\phi}\delta^{(d)}(\mathbf{x} - \mathbf{x'})\dd t, \\
 \left \langle \dd W(\mathbf{x},t)\dd W(\mathbf{x'},t)     \right \rangle & = & 0. 
\end{eqnarray}
{This choice for the noise is based on the physical assumption that its spectral bandwidth is much larger than the characteristic frequency window in which the condensate occurs.
This approximation is almost exact in the coherent pump case where noise is mostly due to the dissipation bath and satisfies $D_{\phi\phi}\approx \gamma$: successful application of SGPE methods to Monte-Carlo studies of the long-range coherence across the optical parametric oscillator (OPO) transition point were reported in~\cite{CARU2005,WOUT2006,RMP2012} but a much more complicate form of the field equation had to be used.

A first attempt to apply stochastic methods to the present case of an incoherent pump was made in~\cite{SAVONA2009}. A more rigorous derivation of the SGPE based on a quantum description of the amplifying mechanism will be published in a future work based on~\cite{ACT_master}. In addition to the unavoidable quantum noise due to the amplification and loss baths, one may however expect that extra classical noise due to pump fluctuations may be playing a dominant role in experiments: also in this case, it appears legitimate to assume it to have a very broad, quasi-white spectrum. 

Our calculations will explicitly show that the long-distance spatial coherence does not depend on the details of the noise spectrum, but only on its intensity at the condensate frequency. The beneficial effect of a frequency dependence of the (deterministic) amplification term in regularizing the UV behavior of the model will finally be discussed in the last part of the Letter.}

\section{Mean-field approximation and collective excitations}
The first step to understand the behavior of the system is to perform a mean-field approximation and look for the stationary state of the deterministic GPE that is obtained by neglecting the noise terms in \eq{GGPE}. For simplicity, we restrict our attention {to spatially homogeneous systems, for which the mean field solution has the translationally invariant form} $\phi(\mathbf{x},t) = \phi_0\,\mathrm{e}^{-i\omega t}$.

 \begin{figure}[h]
 \includegraphics[width=8.8cm]{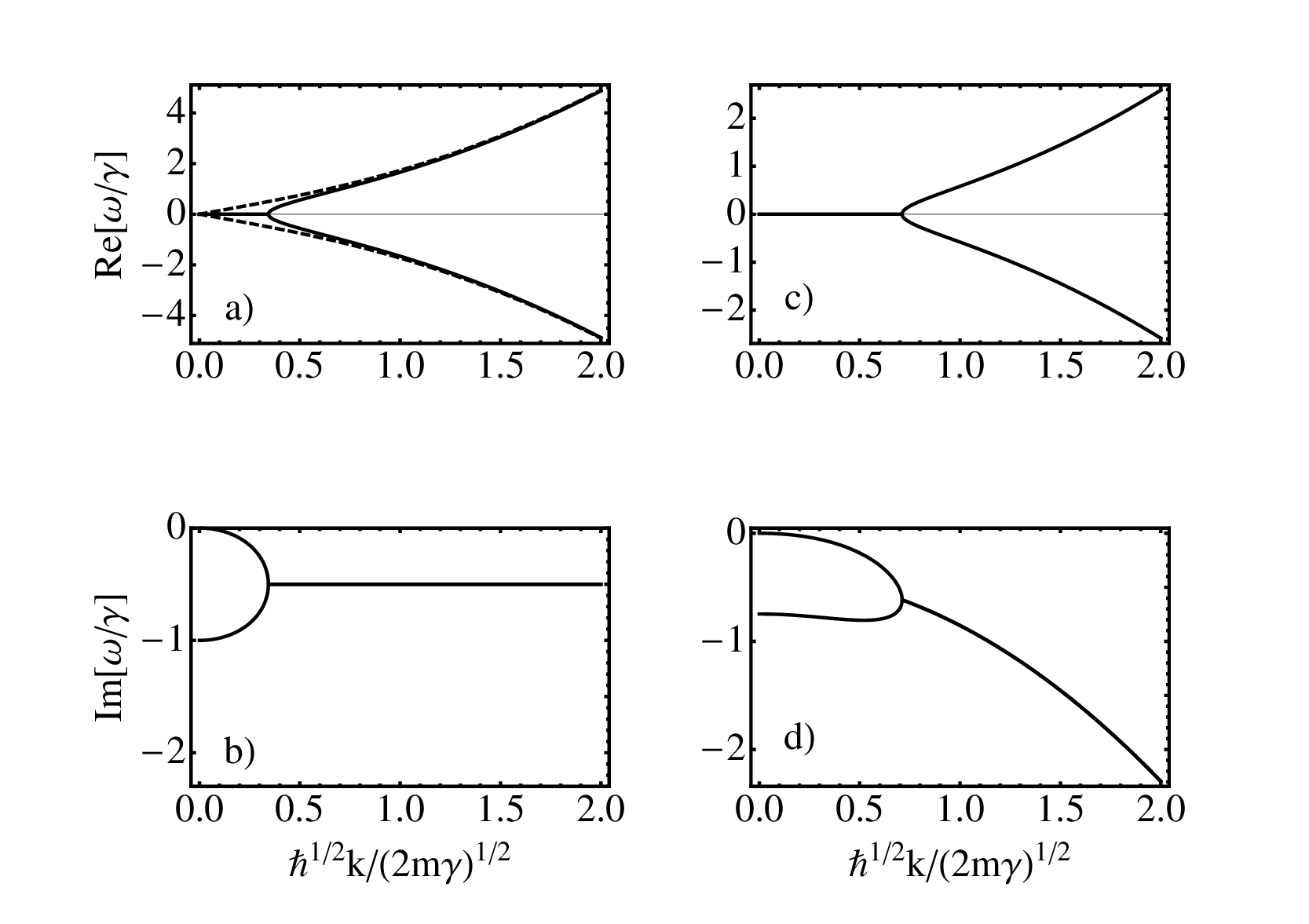}
 \caption{Bogoliubov dispersion of collective excitations of the non-equilibrium condensate as a function of the (real) in-plane momentum $\kk$. Upper panels show the real part of the frequency, lower panels show the imaginary part. The panels (a-b) in the left column are for the case of a frequency independent bath. The panels (c-d) in the right column model the frequency-dependence of the bath as discussed in the last part of the Letter. System parameters: $P_0 = 2\gamma$, $gn_s = \gamma$, $\omega_c = 2\gamma, \omega_0 = 0$. } 
 \label{fig}
 \end{figure}

Independently of the value of the pump intensity $P_0$, the trivial field $\phi=0$ is always a solution of the GPE. For low pump intensities below the threshold $P_0<\gamma$, this is the only solution and is dynamically stable. On the other hand, for pump intensities $P_0>\gamma$ above the threshold, it becomes dynamically unstable and is replaced by another solution with a finite density $|\phi_0|^2  =  n_s( P_0 - \gamma )/\gamma $ and an oscillating frequency $\omega  = \omega_0 + g|\phi_0|^2 $: the global phase of $\phi_0$ acquires a well-defined value by spontaneously breaking the $U(1)$ symmetry of \eq{GGPE}. In the analogy with second order phase transitions, the order parameter hallmarking the transition is then the complex field amplitude $\phi_0$.

The linearization of \eq{GGPE} around the finite solution determines a pair of coupled Bogoliubov partial differential equations for the fluctuation field $\delta\phi(\xx,t)$ defined as $\phi(\xx,t) = [\phi_0 + \delta\phi(\xx,t)]\,\ee^{-i\omega t}$, and its complex conjugate $\delta\phi^*(\xx,t)$. These fluctuations describe the excitations of the Bose field on top of a pure condensate solution, that is the non-condensed fraction of the gas. 

In the spatially homogeneous case under investigation here, it is useful to work in Fourier space and separate the evolution of the different in-plane components. 
For each (real-valued) $\kk$ we are then left with a pair of coupled linear differential equation for $\delta\tilde{\phi}(\kk)$ and $[\delta\tilde{\phi}(-\kk)]^*$, 
\begin{equation}
i\frac{\dd}{\dd t}
\left(
\begin{array}{c}
\delta\tilde{\phi}(\kk) \\ 
(\delta\tilde{\phi}(-\kk))^* 
\end{array}
\right)=\mathcal{L}_\kk
\left(
\begin{array}{c}
\delta\tilde{\phi}(\kk) \\ 
(\delta\tilde{\phi}(-\kk))^* 
\end{array}
\right)
\eqname{Bogo_mot}
\end{equation}
with the $\kk$-dependent Bogoliubov matrix $\mathcal{L}_{\kk}$ defined as 
\begin{equation}
\eqname{Bogo}
\mathcal{L}_\kk=
\left(
\begin{array}{cc}
\epsilon_\kk + \mu -i\Gamma & \mu -i\Gamma\\
-\mu -i\Gamma& -\epsilon_\kk-\mu -i\Gamma
\end{array}
\right)
\end{equation}
in terms of the free particle dispersion $\epsilon_\kk = {\hbar k^2}/{2m}$ and the interaction energy $\mu=g\,|\phi_0|^2$.
By simple diagonalization of $\mathcal{L}_\kk$, one gets a double-branched excitation {spectrum
$\hbar \omega^{\pm}_\kk = -i\hbar \Gamma \pm \hbar\sqrt{E_\kk^2 - \Gamma^2}$,
}
where $E_\kk = \sqrt{\epsilon_\kk(\epsilon_\kk +2\mu)}$ is the standard Bogoliubov dispersion of equilibrium systems. {Remarkably, the effective damping rate $\Gamma= \gamma(P_0-\gamma)/P_0$ vanishes when the critical point is approached from above, while it tends to the bare dissipation rate $\gamma$ far above the threshold.}
An example of this non-equilibrium Bogoliubov dispersion is shown in the left panels of Fig.\ref{fig}: as first noted in~\cite{WOUT2006,SZYM2006,WOUT2007}, its most remarkable feature is the diffusive nature of the Goldstone mode describing {long-wavelength twists of the condensate phase}

\section{Bogoliubov theory of linearized fluctuations}

The linearized field equations \eq{Bogo} are the starting point to study the {small} fluctuations of the Bose field around the pure condensate solution. As usual, this approach is {only accurate far away from the critical point. A complete description of the large fluctuations in the critical region requires more sophisticated approaches and will be postponed to further work. A first remarkable attempt in this direction appeared very recently in~\cite{DIEHL2013}.}
 
{Including the noise term into the RHS of the Bogoliubov equations \eq{Bogo_mot}} one obtains for each $\kk$ a pair of linear Ito stochastic differential equations from which it is straightforward to compute all the steady-state correlations of the field, e.g. the momentum distribution of the particles
\begin{equation}
n_\kk^{ss} = (2\pi)^d\,|\phi_0|^2\,\delta^{(d)}(\kk)\,+\frac{D_{\phi\phi}}{\Gamma} \left[\frac{\mu^2 + \Gamma^2}{E_\kk^2}+ 1 \right];
\label{nk}
\end{equation}
in experiments, this quantity is directly accessible as the angular distribution of the far field emission. The former term in \eq{nk} describes the condensate, while the latter one account for the non-condensed cloud.

{As we shall see in the final part of the Letter, the UV divergence due to the {$\kk$-independent} term in the bracket is an artifact of the model and disappears as soon as the frequency dependence of the bath in included in the theory. On the other hand, the simple model with a spectrally white noise already provides an accurate description of the low-$\kk$, long distance physics: from the frequency-space reformulation of the stochastic Bogoliubov equations, it is in fact immediate to see, for a smooth enough noise spectrum, that the only relevant noise components are the ones in the close vicinity of the condensate frequency $\omega$.} 

\begin{table}[hct]
\centering
\begin{tabular}{| c | c | c | c|}
 \hline
          & Equil. $T = 0$     & Equil. $T \ne 0$ & Non-Equil. \\
 \hline 
$g\ne0$ & $\frac{\sqrt{m\mu}}{2{\hbar}k}$      &  $\frac{m{\kappa_B}T}{{\hbar^2}k^2}$     & $\frac{mD_{\phi\phi}(\mu^2 + \Gamma^2)}{{\hbar}\mu\Gamma}\frac{1}{k^2}$ \\[8pt]
$g=0$   & - & $\frac{2m{\kappa_B}T}{{\hbar^2}k^2}$ & $\frac{4{m^2}D_{\phi\phi}\Gamma}{{\hbar^2}} \frac{1}{k^4}$ \\[5pt]
\hline
\end{tabular}
\caption{{Summary of the low-$\kk$ behaviour of the momentum distribution in the different cases.}}
\label{tab1}
\end{table}

The low-$\kk$ behavior of $n_\kk^{ss}$ in the different cases is summarized in {Table \ref{tab1}} and compared to the prediction of the standard Bogoliubov theory~\cite{PITA2004} for equilibrium systems.
In the presence of interactions $g>0$, the behavior of the non-equilibrium system recovers the one of the corresponding equilibrium system at an effective temperature
\begin{equation}\label{Teff}
 {\kappa_B}T_{\mathit{eff}} = \frac{{\hbar}D_{\phi\phi}(\mu^2 + \Gamma^2)}{\mu\Gamma}:
\end{equation}
independently of the absence of thermalizing collisions between different Bogoliubov modes, the momentum distribution recovers the typical shape of a gas at thermal equilibrium, $n_\kk^{ss} \propto k^{-2}$. As a result, the mere observation of a thermal-like momentum distribution does not appear to be an unambiguous proof of a thermal equilibrium state in the polariton gas. The situation is very different in the case of a non-interacting gas with $g=0$, where the momentum distribution scales as $n_\kk^{ss} \propto k^{-4}$.

{The one-time spatial correlation function is then immediately obtained as the Fourier transform of the momentum distribution $n_\kk^{ss}$,
\begin{equation}
G(\xx - \yy) = \langle\phi^*(\xx)\phi(\yy) \rangle=\int\!\frac{d^d\kk}{(2\pi)^d}\,n_\kk^{ss}\,e^{i\kk(\xx-\yy)}
\end{equation}
According to the Penrose-Onsager criterion, long-range order is defined  by the condition
$ \lim_{|\xx-\yy| \to \infty} G(\xx-\yy) \neq 0$,
which corresponds to the convergence of the momentum distribution $n_\kk^{ss}$ in the IR limit. Using the explicit formula \eq{nk}, one immediately sees that the long-range order in a non-interacting gas requires $d\geq 5$, while for the interacting gase one recovers the equilibrium gas result $d\geq 3$: this result can be seen as a non-equilibrium version of the Mermin-Wagner theorem of equilibrium statistical mechanics~\cite{MERM1966}.}

\section{Density-phase Bogoliubov theory}

{In low-$d$ where the condensate is destabilized by fluctuations, a more sophisticated version of the Bogoliubov theory must be used~\cite{POPO1972,CAST2003}: the Bose field $\phi(\xx,t)$ is expanded in terms of its density and phase according to} 
\begin{equation}
\phi(\mathbf{x},t)  =  \sqrt{n_0 + \delta n(\mathbf{x},t)}\,\mathrm{e}^{i\theta(\mathbf{x},t)}\,\mathrm{e}^{ -i \omega t}.
\end{equation}
 We then assume that density fluctuations $\delta n$ around the average value $n_0$ are weak $\delta n \ll n_0$ and that the phase $\theta$ smoothly varies in space. As the phase $\theta$ can take arbitrarily large values, no assumption on the presence of a condensate is made and the theory is able to describe configurations with no long-range order{, the so-called {\em quasi}-condensates.}

Within this density-phase Bogoliubov approach, the different $\kk$ components of density and phase fluctuations  decouple and evolve according to linear differential equations very similar to \eq{Bogo} including Ito noise terms. 
Averaging over noise, we obtain the following forms for the correlations of density and phase fluctuations at a given $\kk$,
{
\begin{eqnarray}
 \langle \delta n^*_\kk \delta n_{\kk'}\rangle  =  \delta_{\kk\kk'}\frac{2n_0D_{\phi\phi}}{\Gamma}\left(1 -\frac{\mu}{\epsilon_\kk + 2\mu} \right) \label {Ncorr} \\
 \langle \theta^*_\kk\theta_{\kk'}\rangle =  \delta_{\kk\kk'}\frac{D_{\phi\phi}}{2n_0\Gamma}\left(1 + \frac{2(\mu^2 + \Gamma^2)}{E_\kk^2} + \frac{\mu}{\epsilon_\kk + 2\mu}\right) \label{thetacorr} \\
 \langle \delta n^*_\kk \theta_{\kk'} \rangle = \delta_{\kk\kk'}\frac{D_{\phi\phi}}{\epsilon_\kk+2\mu}
\end{eqnarray}}
where $\delta_{\kk\kk'}=(2\pi)^d\delta^{(d)}(\kk-\kk')$.

\section{Spatial coherence in reduced $d$}
Fourier transform of \eq{Ncorr} back to real space shows that the correlation function of density fluctuations quickly tends to zero on a length scale set by the usual healing length $\xi={\sqrt{\hbar/4\mu m}}$. At larger distances, the field correlation functions can then be written in terms of phase fluctuations only. As these inherit the Gaussian statistics of the noise, we can rewrite the field correlation {as
$G(\xx-\yy) \simeq n_0\,\exp[-G_{\theta\theta}(\mathbf{x} - \mathbf{y})/2]$,
}
in terms of the phase correlation {function
$G_{\theta\theta}(\xx-\yy) =\langle \left[\theta(\mathbf{x})-\theta(\mathbf{y})\right]^2 \rangle$.
}

Focussing on the long-distance behavior of the spatial correlation function, we can safely neglect the short range correlations due the first term in \eq{thetacorr} and write the correlation function in the form
\begin{multline}
 G_{\theta\theta}(x) = - {2}\,\xi^2\,\xi_0^{d-4}(\mu)\,\int\frac{\mathrm{d}^d k}{(2\pi)^d}\frac{\ee^{i\kk\cdot\xx}-1}{k^2} + \\
      + {2}\xi^2\,\xi_0^{d-4}(0)\,\int\frac{\mathrm{d}^d k}{(2\pi)^d}\frac{\ee^{i\kk\cdot\xx}-1}{k^2+\xi^{-2}}
\label{Gthetatheta}
\end{multline}
where the additional length scale $\xi_0$ is defined as
\begin{equation}
 \xi_0(\mu) = \left(\frac{4m^2D_{\phi\phi}(\mu^2+\Gamma^2)}{n_0{\hbar^2}\Gamma}\right)^{\frac{1}{d-4}}.
\end{equation}
In the following of this Section, we shall separately consider the physically relevant $d = 1,2,3$ cases, and highlight the peculiarities of each of them. 

\subsection{d = 3}
As we have already mentioned, in $d=3$ both $G_{\theta\theta}$ and $G$ correlation functions have finite long-distance limits as soon as the interaction strength is finite $g>0$: this signals the existence of long-range order, while the momentum distribution of the non-condensed particles follows the typical $k^{-2}$ law of an equilibrium system at the effective temperature $T_{eff}$ determined by \eq{Teff}. On the other hand,  in the opposite case of a non-interacting gas with $g=0$, taking the $\mu\to 0$ limit leads to a field correlation function
\begin{equation}
G(x) \propto \ee^{-\frac{1}{8\pi}\frac{x}{\xi_0(0)}}
\end{equation}
that exponentially decays at large distances within the characteristic length $\xi_0$. This is, of course, in stark contrast with the existence of a BEC transition in the textbook case of a  $d=3$ ideal Bose gas.

\subsection{d = 2}
In this case, the long-distance behavior of the correlation function can be evaluated making use of known properties of the Bessel functions to have the power-law form
\begin{equation}
 G({\xx-\yy}) \simeq \mathrm{const}\frac{\xi^{\eta}}{|\xx-\yy|^{\eta}},
\end{equation}
with an exponent
\begin{equation}
\eqname{eta}
 \eta = \frac{1}{2\pi}\frac{\xi^2}{\xi_0^2(\mu)} = \frac{mD_{\phi\phi}(\mu^2 + \Gamma^2)}{2\pi{\hbar} n_0\mu \Gamma}.
\end{equation}
This power law decay behaviour is characteristic of a superfluid (but not Bose-Einstein condensed) two-dimensional Bose gas at temperatures below the Berezinskii-Kosterlitz-Thouless {(BKT)} transition~\cite{MINN_RMP}. This result was first discussed in a non-equilibrium context in~\cite{SZYM2006} {and applied in a more extensive way to the case of an interacting polariton gas in~\cite{ROUM2012}, where the exponent $\eta$ was explicitely computed}. 
Identifying the superfluid density $n_s$ with the average density $n_0$ and the temperature $T$ with the effective temperature $T_{\mathit eff}$ defined in \eq{Teff}, the form \eq{eta} of the exponent $\eta$ matches the usual form
$\eta_{BKT} = {1}/{\rho_s\lambda_T^2}$,
for equilibrium BKT gases, with the de Broglie wavelength defined as usual as $\lambda_T = [{2\pi\hbar^2/mk_BT}]^{1/2}$. 
{However, while thermodynamic stability of the quasi-ordered BKT phase in equilibrium systems imposes an upper bound $\eta < 1/4$ to the exponent~\cite{MINN_RMP}, no such condition is known for the non-equilibrium case and larger exponents might be observable. This is indeed the case of the experiment~\cite{ROUM2012} where the exponent $\eta$ was measured to be of order $1$.

On the other hand, the field correlation function of a non-interacting $g\to 0$ gas has the form
\begin{equation}\label{quasigauss}
 G_{\theta\theta}(r) \simeq \frac{1}{{4}\pi}\left[1-\mathcal{G} + \ln{2} - \ln\left(\frac{r}{\xi}\right)\right]\left(\frac{r}{\xi_0(0)}\right)^2,
\end{equation}
{where $\mathcal{G}$ is here the Euler-Mascheroni constant, $\mathcal{G}\simeq 0.577 $. As the $g\to 0$ limit corresponds to a diverging value of the healing length $\xi$, the overall trend of $G_{\theta\theta}(r)$ in the $\xi\gg\, r\gg\xi_0$ range can be well approximated by a quadratic growth as a function of $r$. As a result, all the quasi-order that was present in the BKT state disappears in the $g\to 0$ limit and is replaced by a much faster almost Gaussian decay: experimental results shown in the same article~\cite{ROUM2012} for the very weakly interacting photon regime suggest that in this case the correlation function indeed decays in a quasi-Gaussian way, that could be described by \eq{quasigauss}.}

\subsection{d = 1}

Straightforward integration of \eq{Gthetatheta} in the $d=1$ case predicts a fast, diverging growth of $G_{\theta\theta}$ of the form
\begin{equation}
{G_{\theta\theta}(x) = \left[\left(\frac{\xi}{\xi_0(\mu)}\right)^3\frac{x}{\xi} + \left(\frac{\xi}{\xi_0(0)}\right)^3(\ee^{-|x|/\xi}-1) \right].}
\end{equation}
Depending on the relative value of the two characteristic distances $\xi$ and $\xi_0$, this leads to different functional forms of the long-distance tail of $G(x)$.

For $\xi_0 \gg \xi$ (which is the case of an interacting gas at strong pump values well above threshold), the field correlation function has an exponential tail {$G(x) \simeq \exp[-|x|/\ell_1]$} as first predicted in~\cite{WOUT2006} for the OPO pumping scheme. This exponential form closely follow the corresponding case of a $d=1$ gas at equilibrium at a finite $T>0$~\cite{CAST2004}. Also the characteristic length $\ell_1$ has the same form in the two cases: the explicit expression 
\begin{equation}
 \ell_1 = \frac{2n_0{\hbar}\mu\Gamma}{mD_{\phi\phi}(\mu^2 + \Gamma^2)}
\end{equation}
matches the equilibrium one 
 $l^{1D}_T = {n\lambda^2_T}/{\pi}$ 
upon identifying the temperature $T$ with the effective temperature $T_{\mathit{eff}}$ defined in \eq{Teff}. The situation is somehow different in a non-interacting gas and, more in general, for $\xi_0 \ll \xi$: in this case, the long-distance tail can be approximated with a Gaussian form {$G(x) \simeq \exp[-(x/\ell_2)^2]$} of characteristic length $\ell_2 = 2\sqrt{\xi_0^3(0)/\xi}$. Experimental indications of a reduced coherence in one-dimensional polariton condensates was reported in~\cite{CERDA2010,SPAN2012}.

\section{{Frequency-dependent amplification}}
\label{sec:freq-dep}

All the discussion so far has been based on a model assuming amplification and dissipation baths with no frequency dependence. 
{While we expect the white noise approximation to be already a quite accurate description of experiments, both the kinetic models~\cite{PORRAS2002,SARC2008} and the experimental observations of condensation into the ground state of harmonic traps~\cite{SNOK2007} strongly suggest the existence of some deterministic mechanism reinforcing the amplification into the low-energy modes while quenching the one into the high-energy ones.
}

Following~\cite{WOUT2010}, a simple way to model this behavior is to include some frequency-dependence of the amplification on a characteristic scale $\omega_c$, e.g. by replacing the amplification coefficient $P_0$ with the differential operator $P = P_0\,[1 +\omega_c^{-1} (\omega_0 -i\partial/\partial t)]$: in this way, amplification (and then condensation) into low energy modes is favored, while it is completely suppressed above the cut-off frequency $\omega>\omega_c+\omega_0$. As a result, the transition point is still at $P_0^c=\gamma$, but the competition of the frequency-dependent amplification with the interaction-induced blue shift makes the condensate density to saturate at large pump powers, 
$|\phi|^2 = n_s\,(P_0 - P_0^c)/(P_0^c+P_0 g n_s /\omega_c)$.

An example of {the} Bogoliubov excitation spectrum {on top of a spatially homogeneous condensate} is shown in the right panels of Fig.\ref{fig}: while the Goldstone mode with a diffusive dispersion at low $\kk$'s is still visible, the damping rate of the Bogoliubov modes rapidly grows at large $\kk$'s. The momentum distribution of the excitations is {again} obtained by adding noise terms to the Bogoliubov equations. This leads to the closed expression:
\begin{equation}\label{newdistro}
 n_\kk^{ss} = \frac{D_{\phi\phi}}{\widetilde{\Gamma} + M(\epsilon_\kk + \mu)}\left(1 + \frac{\mu^2 + \widetilde{\Gamma}^2}{E_\kk^2}\right),
\end{equation}
in terms of the parameters $\widetilde{\Gamma} = \gamma (P_0 - P_0^c)/[P_0 (1 + gn_s/\omega_c)]$ and $M = P_0/[\omega_c (1 + |\phi_0|^2 /n_s)]$. 
While the low-$\kk$ part of the momentum distribution \eq{newdistro} retains the same behavior as in \eq{nk}, the frequency-dependent amplification is able to regularize the UV behavior of the theory, making all relevant integrals to nicely converge at large $\kk$. 

Of course a quantitative account of the details of the momentum distribution requires a more sophisticated model of the frequency-dependence, still our model is able to qualitatively recover its fast decrease at large $\kk$ as typically observed in Bose-Einstein condensation experiments. In particular, note how our theory predicts that this behavior is completely independent of the collisional thermalization rate: this may explain the unexpected experimental observation of a thermal-like distribution in a gas of non-interacting photons in a VCSEL device in the weak-coupling regime~\cite{BAJO2007}. The {relative impact} of {the frequency-dependent amplification and the noise spectrum on} the thermal photon distribution observed in the photon BEC experiment of~\cite{KLAERS2010} is {presently} under investigation.

\section{Conclusions}

{In this Letter we have introduced a generic phenomenological model of fluctuations on top of a non-equilibrium Bose-Einstein (quasi-)condensate based on a stochastic Gross-Pitaevskii equation. Even though the non-equilibrium stationary state is determined by a dynamical balance of pumping and dissipation rather than by a standard thermodynamical equilibrium condition, {the prediction of the linearized Bogoliubov theory of small fluctuations for the momentum distribution and the long-distance spatial coherence of a weakly interacting gas recovers the one of the corresponding equilibrium system at a finite temperature.} In full agreement with previous works on the subject~\cite{CARU2005,SZYM2006,WOUT2006,SZYM2012,ROUM2012}, this conclusion does not appear to depend on the collisional thermalization rate in the gas nor on the microscopic details of the noise spectrum. 
Thanks to the generic nature of our model, this conclusion applies to a broad range of systems displaying a non-equilibrium BEC-like phase transition, from laser operation in VCSEL devices, to photon BECs, to polariton BECs. }

IC acknowledges partial financial support from ERC through the QGBE grant {and from
Provincia Autonoma di Trento}. We are grateful to Michiel Wouters and Andrea Gambassi for continuous stimulating discussions.

\bibliographystyle{eplbib}
\bibliography{biblio}

\end{document}